# Nonlinear Nonreciprocal Photocurrents under Phonon Dressing

Haowei Xu[1], Hua Wang[1], and Ju Li[1,2, *]

[1] Department of Nuclear Science and Engineering, Massachusetts Institute of Technology, Cambridge, Massachusetts 02139, USA

[2] Department of Materials Science and Engineering, Massachusetts Institute of Technology, Cambridge, Massachusetts 02139, USA

* Corresponding author, liju@mit.edu

## Abstract

Nonlinear optical (NLO) effects have attracted great interest recently. However, by far the computational studies on NLO use the independent particle approximation and ignore many-body effects. Here we develop a generic Green's function framework to calculate the NLO response functions, which can incorporate various many-body interactions. We focus on the electron-phonon coupling and reveal that phonon dressing can make significant impacts on nonlinear photocurrent, such as the bulk photovoltaic (BPV) and bulk spin photovoltaic (BSPV) currents. BPV/BSPV should be zero for centrosymmetric crystals, but when phonons are driven out-of-equilibrium, by for example, a temperature gradient $\nabla T$, the optical selections rules are altered and phonon-pumped BPV/BSPV currents can be non-zero in nominally centrosymmetric crystal. Moreover, we elucidate that such NLO responses under non-equilibrium phonon dressing can be nonreciprocal, as the direction of the current does not necessarily get reversed when the direction of the temperature gradient is reversed.

**Introduction**. Light is a valuable tool for characterizing diverse material properties, including topological quantities such as the Berry curvature [1–8]. Light can also drive electronic, phononic and nuclear-spin excitations [9,10] and even trigger ionic [11–13] and electronic [14–16] phase transitions. On the other hand, light-matter interactions can be used to detect and generate light. For example, the spontaneous parametric down-conversion in nonlinear optical (NLO) materials can be applied to create entangled photon pairs used in quantum information technologies [17].

To better understand and harness light-matter interactions, it is crucial to obtain the response functions from first principles. Most works hitherto rely on the independent particle approximation (IPA), which assumes that electrons are (nearly) free particles and ignores many-body interactions. This is unsatisfactory since many-body interactions can play a central role in quantum materials [18]. For example, excitonic effects can remarkably enhance NLO responses [19–21]. Even for conventional materials, it is indispensable to include many-body interactions such as electron-phonon (e-ph) coupling [22], if one aims to accurately calculate the response functions. Therefore, it is highly desirable to have a theoretical framework to systematically incorporate many-body interactions when calculating the response functions.

In this work, we formulate a generic Green's function framework, which can be used to deal with various many-body interactions. We use e-ph coupling as an example to demonstrate how many-body interactions can make quantitative and even qualitative impacts on light-matter interactions. We will focus on nonlinear photocurrents [23–25] such as the bulk photovoltaic (BPV) and bulk spin photovoltaic (BSPV) effects, while the methodologies in this work also apply to other NLO responses, including higher-order NLO effects. We start with the temperature-gradient effect on BPV responses. If the phonons are at equilibrium, the IPA can give reasonably good results compared with the many-body formalism. In contrast, when phonons are out-of-equilibrium, rich physics can arise, which cannot be captured if e-ph couplings are ignored. Using a two-temperature model, we illustrate that phonon dressing can effectively break the original spatial symmetries and consequently alter the optical selection rules. We provide a geometrical interpretation for this effect. We also reveal that the nonlinear photocurrent under phonon dressing can be nonreciprocal, in that for nominally centrosymmetric crystals, the direction of the current does not necessarily get reversed when the direction of the temperature gradient is reversed. We also introduce a symmetry-breaking vector, which quantifies and unifies the symmetry breaking from both atomic structures (intrinsic) and phonon dressings (extrinsic).

**Green's Function Framework**. First, we briefly introduce the Green's function framework and demonstrate how many-body interactions can be incorporated. The BPV effect can be described as

$$j^a = \sigma^a_{bc}(\omega)\mathcal{E}^b(\omega)\mathcal{E}^c(-\omega) \quad (1)$$

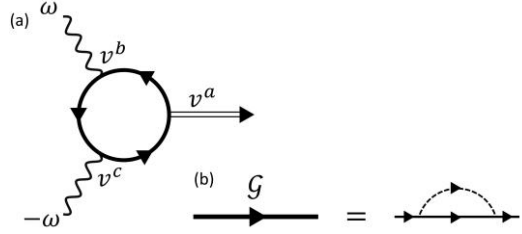

**Figure 1 (a)** Feynman diagram of the BPV process. **(b)** Electron propagator under phonon dressing. Think (thin) solid arrow: dressed (undressed) electron propagator. Wavy line: photon propagator. Dashed arrow: phonon propagator. Double arrow: charge current output.

Here $\mathcal{E}(\omega)$ is the alternating electric field with frequency $\omega$. $j^a$ is the DC charge current. $a$ and $b/c$ denote the directions of the current and the electric field, respectively. $\sigma_{bc}^a(\omega)$ is the BPV conductivity and can be expressed as

$$\sigma_{bc}^a(\omega) = -\frac{ie^3}{\omega^2} \int \frac{d\bm{k}}{(2\pi)^d} \int \frac{dE}{2\pi} \text{Tr}\{v^a \tilde{\mathcal{G}}_{bc}^<(E)\} \tag{2}$$

where

$$\tilde{\mathcal{G}}_{bc}^< = \mathcal{G}^r(E) v^b \mathcal{G}^r(E+\omega) v^c \mathcal{G}^<(E) + \mathcal{G}^r(E) v^b \mathcal{G}^<(E+\omega) v^c \mathcal{G}^a(E)$$
$$+ \mathcal{G}^<(E) v^b \mathcal{G}^a(E+\omega) v^c \mathcal{G}^a(E) + (b \leftrightarrow c, +\omega \leftrightarrow -\omega)$$

The Feynman diagram for the process above is shown in Figure 1. For simplicity, we did not include terms that may arise from higher-order band dispersions in a tight-binding model. See Supplementary Materials [26] and Refs. [6,27] for detailed discussions. $(b \leftrightarrow c, +\omega \leftrightarrow -\omega)$ indicates simultaneous exchange between $b, c$ and $+\omega, -\omega$. $e$ is the electron charge, d is the dimension of the system, and $v$ is the electron velocity. $\tilde{\mathcal{G}}$ and $\mathcal{G}$ denote Green's functions with and without light illumination, while the superscripts r, a, and < denote retarded, advanced, and lesser Green's functions, respectively. Both $v$ and $\mathcal{G}/\tilde{\mathcal{G}}$ are functions of $\bm{k}$-point in the first Brillouin zone, and the $\bm{k}$ arguments are occasionally omitted for simplicity. $E$ is a frequency/energy argument (the Planck constant $\hbar$ is occasionally set as 1 in the following). The many-body interactions are incorporated into the Green's function $\mathcal{G}$, which can be obtained perturbatively with e.g., Feynman diagrams or non-perturbatively with other methods [28].

The Migdal electron self-energy due to e-ph coupling can be expressed as [28–30]

$$\Sigma_{nn'}(E, \bm{k}) = \sum_{m, q\nu} g^*_{mn\nu}(\bm{k}, \bm{q}) g_{mn'\nu}(\bm{k}, \bm{q}) \tag{3}$$
$$\times \left[ \frac{n_{q\nu} + 1 - f_{m, \bm{k}+\bm{q}}}{E - \epsilon_{m, \bm{k}+\bm{q}} - \omega_{q\nu} + i\delta_p} + \frac{n_{q\nu} + f_{m, \bm{k}+\bm{q}}}{E - \epsilon_{m, \bm{k}+\bm{q}} + \omega_{q\nu} + i\delta_p} \right]$$

where $g_{mn\nu}(\bm{k}, \bm{q}) \equiv \left(\frac{\hbar}{2m_0 \omega_{q\nu}}\right)^{\frac{1}{2}} \langle \psi_{m, \bm{k}+\bm{q}} | \partial_{q\nu} V | \psi_{n\bm{k}} \rangle$ is electron-phonon coupling matrix, with $m, n$ as electron band indices, $\psi$ as electron wavefunction, $V$ as the self-consistent potential, $\boldsymbol{q}, \nu$ and $\omega_{q\nu}$ as the phonon wavevector, branch, and frequencies, respectively. $\epsilon_{n\boldsymbol{k}}$ is the electronic band energy. $f_{n\boldsymbol{k}}$ and $n_{q\nu}$ are electron and phonon occupation numbers. $m_0$ is a mass parameter that makes $\Sigma$ have the unit of energy. Here we adopt the convention in the EPW [30,31] package and take $m_0$ as the proton mass. Note that if one sticks to the same $m_0$ throughout the calculation process, then the final results would be the same. $\delta_p$ corresponds to phonon lifetime and is set as 50 meV. Varying $\delta_p$ would not change the essence of our general results (see Figure S6 in Ref. [26]). Here we do not include the static Debye-Waller term in the electron self-energy, whose contributions to the BPV conductivities are small according to our tests. The electron Green's function can be obtained from the Dyson's equation [28]

$$\mathcal{G}^{-1}(E, \boldsymbol{k}) = G_0^{-1}(E, \boldsymbol{k}) - \Sigma(E, \boldsymbol{k}) \quad (4)$$

where $G_0(E, \boldsymbol{k}) = \frac{1}{E - H_{\boldsymbol{k}} + i\delta_e}$ is the non-interacting Green's function, with $H_{\boldsymbol{k}}$ as the electron single-particle Hamiltonian. $\delta_e$ represents electron self-energy due to interactions with defects, other electrons, etc. (excluding interactions with phonons). $\delta_e$ is taken as 20 meV (corresponding to relaxation time ~0.2 ps), based on recent experimental results [32,33].

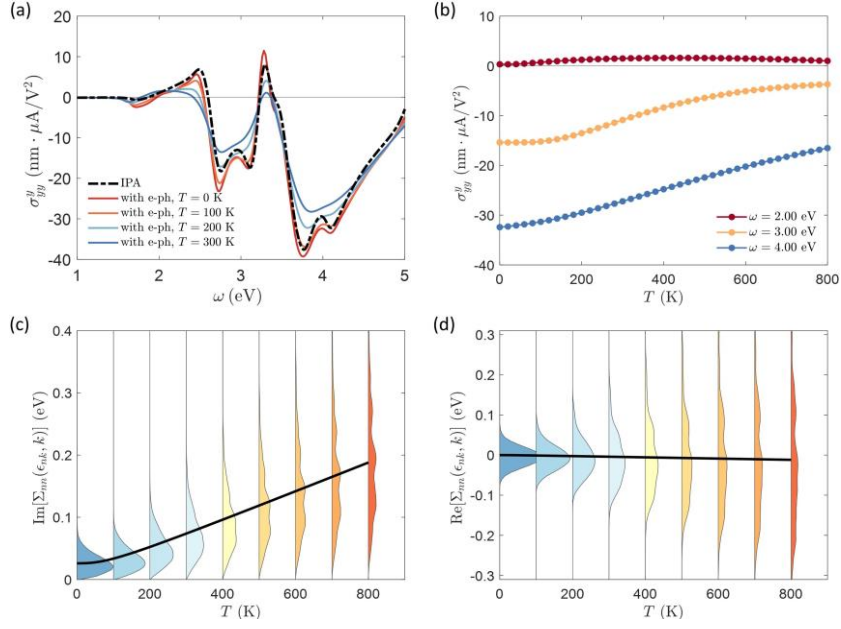

**Figure 2. Phonons in equilibrium.** (a) $\sigma_{yy}^{y}(\omega, T)$ of monolayer WSe$_2$ at selected temperatures. Colored solid curves are from many-body formalism while the black dashed curve is from IPA. (b) $\sigma_{yy}^{y}(\omega, T)$ as a function of $T$ for selected $\omega$. (c) imaginary and (d) real part of electron self-energy due to e-ph coupling. The colored areas represent the distribution of self-energy at selected temperatures (labelled on the $x$-axis), while the black curve is the mean electron self-energy.

In the following, we will use monolayer WSe$_2$ as an example to illustrate the phonon dressing effects on BPV. We use monolayer WSe$_2$ because some concepts can be easily demonstrated and visualized in 2D materials. The main conclusions in this work apply to 3D materials as well. As an example, we also studied 3D silicon, which has inversion symmetry. It is shown that temperature difference can effectively break the inversion symmetry and produce a DC photocurrent in silicon (Figure S5 in Ref. [26]). We will focus on the BPV charge photocurrents under linearly polarized light, while similar conclusions should also carry over to circularly polarized light, and spin current under BSPV as well [34].

**Temperature Effect on BPV: Phonons in Equilibrium.** First, we consider a phonon ensemble with equilibrium occupation $n_{qv} = n_{\text{BE}}(\omega_{qv}, T)$, with $n_{\text{BE}}$ as the Bose-Einstein distribution and $T$ as the temperature. The BPV conductivities $\sigma_{bc}^a(\omega, T)$ is a function of $T$, due to the temperature-dependence of electron self-energies in Eq. (3). $\sigma_{yy}^y(\omega, T)$ of WSe$_2$ at several temperatures are shown in Figure 2a. One can see that at elevated temperatures, some resonance peaks in the $\sigma_{yy}^y(\omega, T)$ vs. $\omega$ spectrum disappear, and the magnitude of $\sigma_{yy}^y(\omega, T)$ decreases with $T$ in a sublinear fashion (Figure 2b). This is because phonon populations increase with temperature, leading to faster e-ph scatterings and thus shorter electron lifetimes. The distributions of the imaginary and real part of the phonon self-energy at band energies $\Sigma_{nn}(\epsilon_{nk}, \boldsymbol{k})$ are shown in Figures 2c, 2d (filled areas), respectively. Generally, $\Sigma_{nn}(\epsilon_{nk}, \boldsymbol{k})$ has larger absolute values and wider distributions at higher temperatures. The mean electron self-energies defined as $\langle \Sigma \rangle \equiv \frac{1}{N} \sum_{nk} \Sigma_{nn}(\epsilon_{nk}, \boldsymbol{k})$ with $N$ as the number of total electron modes, are shown as black curves in Figures 2c, 2d. Im[$\langle \Sigma \rangle$] grows linearly with $T$ for $T \gtrsim 100$ K, while Re[$\langle \Sigma \rangle$] is close to 0 and is almost independent of $T$.

The calculations above use the many-body formalism and explicitly incorporate e-ph coupling. The BPV conductivities can also be calculated using IPA. In this case, the e-ph coupling effect can be approximated with a mode-independent relaxation time $\langle \tau \rangle = \frac{\hbar}{\text{Im}[\langle \Sigma \rangle]}$. Here we take $\hbar/\langle \tau \rangle = 26$ meV, corresponding to the mean self-energy at $T = 0$ (Figure 2c). $\sigma_{yy}^y(\omega)$ based on IPA is shown as the black dashed curve in Figure 2a, which in general agrees with the $\sigma_{yy}^y(\omega, T = 0$ K) from the many-body formalism. We also find similar agreements for other components of the $\sigma_{bc}^a$ tensor at other temperatures. This indicates that when phonons are in thermal equilibrium, IPA with mode-independent relaxation time can give reasonably good results. However, this is not the case when phonons are out-of-equilibrium, as we will show below.

**Symmetry Breaking from Temperature Difference.** In the previous section, we considered an equilibrium phonon distribution under uniform temperature. Richer physics can arise when the phonons are out-of-equilibrium, driven by the boundary condition or volumetric pumping by e.g. another photon source. Typically, phonons are out-of-equilibrium when there is a temperature difference – if the temperature at the left boundary $T_1$ is higher than that at the right boundary $T_2$, there will be more phonons traveling to the right. Here we adopt a simplified two-temperature model to describe non-equilibrium phonons – phonons with velocities $\boldsymbol{v}_{qv} \cdot \widehat{\boldsymbol{x}} > 0$ have occupations number $n_{qv}^1 = n_{\text{BE}}(\omega_{qv}, T_1)$, while those with $\boldsymbol{v}_{qv} \cdot \widehat{\boldsymbol{x}} < 0$ have $n_{qv}^2 = n_{\text{BE}}(\omega_{qv}, T_2)$. $\boldsymbol{v}_{qv}$ is the group velocity of the phonon and $\widehat{\boldsymbol{x}}$ is the unit vector along transport $x$-direction (Figure 3c). The phonon occupations $n_{qv}^1$ and $n_{qv}^2$ are inserted into Eq. (3), yielding the electron self-energy, which is in turn used to calculate the BPV conductivities $\sigma_{bc}^a(\omega, T_1, T_2)$. Here we have made two approximations: 1) the phonon occupations obey an ideal two-temperature model, and 2) we extend the Migdal self-energy to non-equilibrium systems. These two approximations are reasonable when the system is slightly out-of-equilibrium ($|T_1 - T_2| \ll T_1$). However, to reduce numerical noise, $|T_1 - T_2| \sim T_1$ is used. The essence of the results should be qualitatively valid. Quantitatively, the phonon dressing effects can be weaker in reality, since the phonon distributions tend to thermalize and should be less extreme than the two-temperature model described above. To rigorously calculate the BPV responses under a continuous temperature gradient, one may solve the Boltzmann transport equation that couples the electron and phonon distributions. But this would greatly complicate both theoretical and computational analyses, and we would like to leave this for future works.

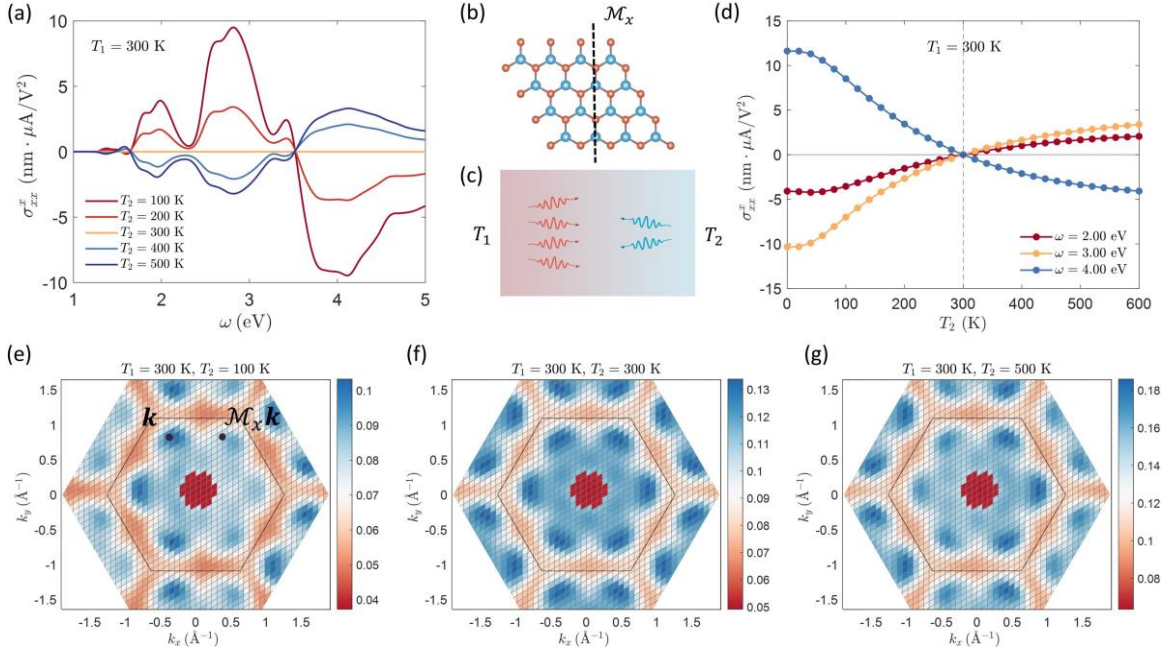

**Figure 3. Two-temperature model.** (a) $\sigma_{xx}^{x}(\omega, T_1, T_2)$ of monolayer WSe$_2$ in the two-temperature model with $T_1 = 300$ K and varied $T_2$. (b) atomic structure of monolayer WSe$_2$ with $\mathcal{M}_x$ symmetry labelled by the dashed line. (c) Illustration of the two-temperature model. Wavy arrows denote phonons. (d) $\sigma_{xx}^{x}(\omega, T_1, T_2)$ as a function of $T_2$ with selected $\omega$ and $T_1 = 300$ K. (e, f, g) electron self-energy Im$[\Sigma_{nn}(E, \boldsymbol{k})]$ with selected $T_1$ and $T_2$. $n$ corresponds to the lowest conduction band, and $E$ is 0.2 eV higher than the conduction band minimum. The black hexagons denote the first Brillouin zone. The color-bars are in the unit of eV.

A prominent effect of the temperature gradient is symmetry breaking. Due to mirror symmetry $\mathcal{M}_x$ (Figure 3b) of monolayer WSe$_2$, some elements of the BPV tensors, such as $\sigma_{xx}^{x}$, are forbidden. When phonons are in equilibrium, $\mathcal{M}_x$ will be preserved even if e-ph couplings are considered. However, when more phonons are traveling to the right ($T_1 > T_2$, Figure 3c), $+x$ and $-x$ will be inequivalent for phonons. The $\mathcal{M}_x$ symmetry breaking can be transmitted to the electronic system due to asymmetry scatterings with asymmetric phonons, leading to nonzero $\sigma_{xx}^{x}$. This effect is illustrated in Figures 3a, 3d, where we fix $T_1 = 300$ K and vary $T_2$. One can see that when $T_2 = T_1 = 300$ K, $\sigma_{xx}^{x}$ is unanimously zero for all light frequencies. In contrast, when $T_2 \neq T_1$, $\sigma_{xx}^{x}$ can be non-zero, and the sign of $\sigma_{xx}^{x}$ is opposite for $T_2 > T_1$ and $T_1 > T_2$. This is different from coherently exciting phonons via THz irradiation [35], which triggers transient symmetry breaking from equilibrium.

The $\mathcal{M}_x$ symmetry breaking in the electron system can be directly visualized in the Migdal self-energy. Here we plot $\text{Im}[\Sigma_{nn}(E, \boldsymbol{k})]$ as a function of $\boldsymbol{k}$, with $n$ corresponding to the lowest conduction band, and $E$ fixed as 0.2 eV higher than the conduction band minimum. When $T_1 = T_2 = 300$ K, the $\mathcal{M}_x$ symmetry in $\text{Im}[\Sigma_{nn}(E, \boldsymbol{k})]$ is preserved (Figure 3f). This corroborates that $\mathcal{M}_x$ symmetry in the electron system cannot be broken by phonons in equilibrium. On the other hand, when $T_1 \neq T_2$, the $\mathcal{M}_x$ symmetry in $\text{Im}[\Sigma_{nn}(E, \boldsymbol{k})]$ disappears (Figures 3e, 3g). In other words, electrons with wavevector $\boldsymbol{k}$ are scattered by phonons at different rates than those with $\mathcal{M}_x \boldsymbol{k}$ ($\mathcal{M}_x \hat{\star}$ denotes the mirror-$x$ image of $\hat{\star}$). Under light illumination, contributions from electrons at $\boldsymbol{k}$ and $\mathcal{M}_x \boldsymbol{k}$ would not cancel, resulting in non-zero $\sigma_{xx}^x$.

The symmetry breaking due to phonon dressings can be interpreted geometrically as well. Electrons can be represented by the localized Wannier functions $|m\boldsymbol{R}\rangle$, where $m$ and $\boldsymbol{R}$ label the orbital and the unit-cell, respectively. The rate of phonon-assisted electron jumping in real space is $p_{m\boldsymbol{R} \to m'\boldsymbol{R}'} \propto n_{qv}|\langle m'\boldsymbol{R}'|\partial_{qv} V|m\boldsymbol{R}\rangle|^2$. Due to $\mathcal{M}_x$ symmetry, one has $|\langle m'\boldsymbol{R}'|\partial_{qv} V|m\boldsymbol{R}\rangle|^2 = |\langle m', \mathcal{M}_x \boldsymbol{R}'|\partial_{\mathcal{M}_x q,v} V|m, \mathcal{M}_x \boldsymbol{R}\rangle|^2$. When phonons are in equilibrium, one has $n_{qv} = n_{\mathcal{M}_x q, v}$, and thus $p_{m\boldsymbol{R} \to m'\boldsymbol{R}'} = p_{m, \mathcal{M}_x \boldsymbol{R} \to m', \mathcal{M}_x \boldsymbol{R}'}$. However, if $n_{qv} \neq n_{\mathcal{M}_x q, v}$, then $p_{m\boldsymbol{R} \to m'\boldsymbol{R}'} \neq p_{m, \mathcal{M}_x \boldsymbol{R} \to m', \mathcal{M}_x \boldsymbol{R}'}$, indicating that scatterings between $+x$ and $-x$ are asymmetrical, and that electrons have tendencies to move in a certain direction (either $+x$ or $-x$). Consequently, the free carriers generated by light would have an overall displacement in real space, leading to a net charge current. This geometrical interpretation is similar to that of the shift current [36], where the asymmetrical scatterings come from asymmetrical atomic potentials.

**Nonreciprocal Phonon Dressing.** One interesting observation from Figures 2 and 3 is the large magnitude of $\sigma_{xx}^x$ in the two-temperature model. When $|T_1 - T_2| = 200$ K, the differences between the self-energies of electrons at $\boldsymbol{k}$ and $\mathcal{M}_x \boldsymbol{k}$ are on the order of 0.01 eV (Figures 3e, 3g). This is rather small compared with the typical energy scales involved in NLO responses, such as the electron band energies and the light frequency, both of which are ~1 eV. Remarkably, the small asymmetry in electron self-energies leads to quite a large $\sigma_{xx}^x$. Comparing Figure 3a and Figure 2a, one can see that the magnitude of $\sigma_{xx}^x(\omega, T_1 = 100 \text{ K}, T_2 = 300 \text{ K})$ can reach ~30% of that of $\sigma_{yy}^y(\omega, T = 100 \text{ K or } 300 \text{ K})$. Note that $\sigma_{yy}^y$ in Figure 2a comes from the intrinsic asymmetry in the atomic structure of WSe$_2$. To understand this phenomenon, one can look at the $\boldsymbol{k}$-resolved contributions to the total BPV conductivities, defined as $I_{bc}^a(\boldsymbol{k}) \equiv -\frac{ie^3}{\omega^2} \int \frac{dE}{2\pi} \text{Tr}\{v^a \tilde{\mathcal{G}}_{bc}^<(E)\}$. Here we fix $\omega = 2$ eV and $T_1 = T_2 = 300$ K. $I_{xx}^x(\boldsymbol{k})$ for $\sigma_{xx}^x$ and $I_{yy}^y(\boldsymbol{k})$ for $\sigma_{yy}^y$ are plotted in Figure 4a and 4b, respectively. Due to the $\mathcal{M}_x$ symmetry, one has $I_{xx}^x(\boldsymbol{k}) = -I_{xx}^x(\mathcal{M}_x \boldsymbol{k})$. In contrast, mirror-$y$ $\mathcal{M}_y$ symmetry is broken in WSe$_2$, and hence $I_{yy}^y(\boldsymbol{k}) \neq -I_{yy}^y(\mathcal{M}_y \boldsymbol{k})$. However, the difference between $I_{yy}^y(\boldsymbol{k})$ and $-I_{yy}^y(\mathcal{M}_y \boldsymbol{k})$ is barely noticeable in Figure 4b. Quantitatively, integrating the absolute value of $I_{yy}^y(\boldsymbol{k})$ gives $\left|\int \frac{d\boldsymbol{k}}{(2\pi)^2} I_{yy}^y(\boldsymbol{k})\right| \sim 0.01 \times \int \frac{d\boldsymbol{k}}{(2\pi)^2} |I_{yy}^y(\boldsymbol{k})|$. This suggests that the $\mathcal{M}_y$ symmetry

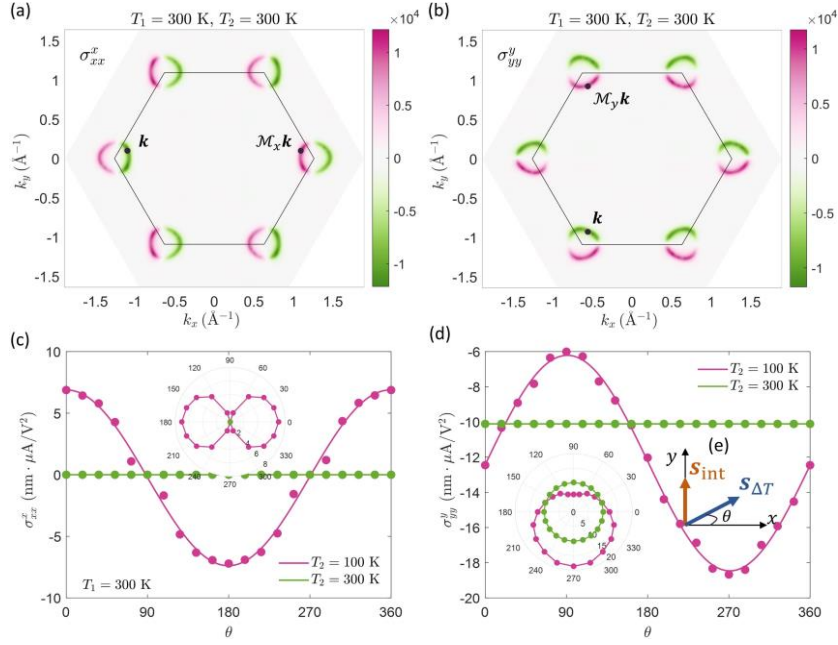

**Figure 4. Nonreciprocal phonon dressing. (a, b)** $\mathbf{k}$-resolved contributions to the BPV conductivities for (a) $\sigma_{xx}^x$ and (b) $\sigma_{yy}^y$ at $\omega = 2$ eV and $T_1 = T_2 = 300$ K. The black hexagon is the first Brillouin zone. The color-bars are in the unit of Å³ ·A/V². **(c)** $\sigma_{xx}^x(\omega, T_1, T_2)$ as a function of $\theta$ with $\omega = 3$ eV, $T_1 = 300$ K. Pink and green curves correspond to $T_2 = 100$ and 300 K, respectively. The green curve is zero for all $\theta$. **(d)** similar to (c), but for $\sigma_{yy}^y$. The insets in c (d) are $\sigma_{xx}^x$ ($\sigma_{yy}^y$) vs. $\theta$ plotted with polar axes. **(e)** An illustration of the symmetry breaking vector.

breaking in $I_{yy}^y(\mathbf{k})$ is only ~ 1%, and that most contributions from $\mathbf{k}$ are canceled by those from $\mathcal{M}_y\mathbf{k}$. As a result, the magnitude of $\sigma_{yy}^y$ is limited.

To further quantify the symmetry break strength, we introduce a symmetry-breaking vector $\mathbf{s}$. Note that $\mathbf{s}$ is a macroscopic quantity. Each electron $m\mathbf{k}$ may have its own symmetry condition (e.g., $\mathbf{k}$ can be a low symmetry point in the Brillouin zone), and $\mathbf{s}$ corresponds to the "average" or "global" symmetry condition of all electrons. The intrinsic asymmetry in the atomic structure of monolayer WSe$_2$ leads to $\mathbf{s}_{\text{int}} = (0, a_0, 0)$ with $a_0$ as a constant, which breaks $\mathcal{M}_y$ and preserves $\mathcal{M}_x$. The temperature difference described above leads to an extrinsic symmetry breaking. Here we generalize the temperature difference to an arbitrary direction. Assuming that the temperature difference is along $\hat{\mathbf{n}} = (\cos\theta, \sin\theta, 0)$, then phonons with $\mathbf{v}_{qv} \cdot \hat{\mathbf{n}} > 0$ and $\mathbf{v}_{qv} \cdot \hat{\mathbf{n}} < 0$ have temperature $T_1$ and $T_2$, respectively. The macroscopic symmetry breaking vector $\mathbf{s}_{\Delta T}$ induced by asymmetric phonons dressing should be parallel to $\hat{\mathbf{n}}$, hence $\mathbf{s}_{\Delta T} = (b\cos\theta, b\sin\theta, 0)$ with $b = b(T_1, T_2)$ as a function of both $T_1$ and $T_2$. The total symmetry breaking vector is (Figure 4e)

$$\mathbf{s} = \mathbf{s}_{\text{int}} + \mathbf{s}_{\Delta T} = (b\cos\theta, a_0 + b\sin\theta, 0) \quad (5)$$

The BPV conductivities $\sigma_{bc}^a$ should be a function of $\mathbf{s}$. Here we fix $\omega = 3$ eV. When $T_1 = T_2$, one has $b = 0$ and $\mathbf{s} = (0, a_0, 0)$, hence both $\sigma_{xx}^x$ and $\sigma_{yy}^y$ are independent of $\theta$ (green curves in Figures 4c, 4d). When $T_1 \neq T_2$ ($T_1 = 300$ K, $T_2 = 100$ K here), one has $b \neq 0$, then $\sigma_{xx}^x$ and $\sigma_{yy}^y$ would be $\theta$-dependent (pink curves in Figures 4c, 4d). For $\sigma_{xx}^x$, $s_x = b\cos\theta$ is relevant, and one has $\sigma_{xx}^x \approx 7.1 \times \cos\theta$ [nm·µA/V$^2$]. As for $\sigma_{yy}^y$, $s_y = a_0 + b\sin\theta$ is relevant, and again one has the sinusoidal relationship $\sigma_{yy}^y \approx -12.3 + 6.1 \times \sin\theta$ [nm·µA/V$^2$]. One can deduce that $\left|\frac{a_0}{b}\right| \sim \frac{12.3}{6.1} \sim 2$, indicating that the magnitude of $\boldsymbol{s}_{\text{int}}$ and $\boldsymbol{s}_{\Delta T}$ are close. As discussed before, the asymmetry in electron self-energies due to asymmetric phonon dressing is weak (~100 times smaller than the typical energy scale in optical processes). Therefore, the intrinsic symmetry breaking is weak, there is plenty of room to enhance the BPV conductivities by breaking the symmetries to a greater extent. Potential ways of breaking symmetries may be strain gradient, electric field, magnetic field, etc.

Interestingly, the BPV current under phonon dressing is nonreciprocal [37,38]. As shown in Figure 4d, the photocurrent does not necessarily change direction when the direction of the temperature difference reverses ($\theta \to \pi + \theta$). This comes from the *interference* between the asymmetry from 1) extrinsic phonon dressing and 2) intrinsic atomic potential [Eq. (5)]. The intrinsic shift current is the consequence of the so-called shift vector $\boldsymbol{R}_{\text{int}}$ – upon photoexcitation, the electron wave package can jump in real space by $\boldsymbol{R}_{\text{int}}$, owing to scatterings with asymmetric atomic potentials. The average $\boldsymbol{R}_{\text{int}}$ (over all electron modes) is closely related to the intrinsic symmetry breaking and should be parallel with $\boldsymbol{s}_{\text{int}}$. The scatterings with asymmetrical phonons, on the other hand, lead to another electron displacement $\boldsymbol{R}_{\Delta T}$ in real space as discussed before, resulting in an additional term in the total current. The average $\boldsymbol{R}_{\Delta T}$ should align in parallel with $\boldsymbol{s}_{\Delta T}$. Consequently, one has the total current $\boldsymbol{j} \propto \boldsymbol{R}_{\text{int}} + \boldsymbol{R}_{\Delta T} \propto \boldsymbol{s}_{\text{int}} + \boldsymbol{s}_{\Delta T}$. When the direction of the temperature difference reverses ($\boldsymbol{s}_{\Delta T} \to -\boldsymbol{s}_{\Delta T}$), $\boldsymbol{j}$ does not necessarily change sign. Note that time-reversal symmetry $\mathcal{T}$ is broken in our two-temperature model – when there is a temperature difference, the system would thermalize, leading to entropy increase and $\mathcal{T}$ breaking. There are also other sources of dissipation in the nonlinear photocurrent generation process. For example, under light illumination, some electrons would be excited to the conduction bands. There will be dissipations when these electrons jump back to the valence bands. The important role of dissipation is discussed in Refs. [39,40].

In summary, we developed a Green's function formalism to systematically incorporate the many-body interactions in NLO processes. We use e-ph coupling as an example and demonstrate that the phonon dressing effect can make significant impacts on NLO responses. Notably, out-of-equilibrium phonons can lead to symmetry breaking in nominally centrosymmetric crystals and alter the selection rules on NLO processes. We also elucidate that the phonon dressing effect can be nonreciprocal. This work paves the way for future studies on the correlations between many-body theory and light-matter interactions.

**Acknowledgments.** This work was supported by an Office of Naval Research MURI through grant #N00014-17-1-2661. Haowei Xu acknowledges helpful discussions with Meihui Liu.